\documentstyle[aps,preprint,epsfig]{revtex}

\begin{document}

\preprint{{\large SUNY-NTG-97-07}}

\title{Strangeness production and flow in heavy-ion collisions}
\bigskip
\author{G. Q. Li$^{1,2}$, G. E. Brown$^1$, C.-H. Lee$^1$, and C. M. Ko$^2$}
\address{$^{1}$ Department of Physics, State University of New
York at Stony Brook, Stony Brook, NY 11794\\
$^{2}$ Cyclotron Institute and Physics Department,
Texas A\&M University, College Station, Texas 77843}

\maketitle
  
\begin{abstract}
We study strangeness ($K^+$, $K^-$, and $\Lambda$)
production and flow in Ni+Ni collisions at 1-2 AGeV, based
on the relativistic transport model including
the strangeness degrees of freedom. We find that strangeness
spectra and flow are sensitive to the properties of strange
hadrons in nuclear medium. The predictions of the chiral
perturbation theory that the $K^+$ feels a weak repulsive potential
and $K^-$ feels a strong attractive potential are in good agreement
with recent experimental data from FOPI and KaoS collaborations. 
\end{abstract}

\section{INTRODUCTION}

With the development of various heavy-ion facilities, 
nuclear physics is expanding into many new directions. One of these
is the study of the properties of strange particles, 
namely hyperons and kaons, in dense matter. Strangeness plays
a special role in the development of hadronic and nuclear
models. The mass of strange quark is about 150 MeV, which is,
on the one hand, considerably larger than the mass of light 
(up and down) quarks ($\sim 5$ MeV), but on the other hand, 
much smaller than that of charm quark ($\sim 1.5$ GeV). 
In the limit of vanishing quark mass the chiral symmetry is good, 
and systematic studies can be carried out using chiral perturbation 
theory for hadrons made of light quarks. On the other hand, 
if the quark mass is large,  one can use the non-relativistic quark model
to study the properties of charmed and heavier hadrons.
The hadrons with strangeness lie in between these two limits and
therefore present a challenge to the theorists.

Nevertheless, chiral perturbation calculations have been extensively
and quite successfully carried out in the recent past for the study 
of kaon-nucleon ($KN$) and antikaon-nucleon (${\bar K}N$) scattering 
{\cite{lee,weise}}. The main reason for the
success of the chiral perturbation theory is the existence of
a large body of experimental data that can used to constrain the
various parameters in the chiral Lagrangian.  

Of great interest to nuclear physicists is the study of
the properties of strange hadrons in dense matter. 
Chiral perturbation calculations in matter are still 
under development, since a new energy scale, the Fermi
momentum, is now involved. For this kind of theoretical
development to be successful, a large body of experimental
data that can be obtained only from heavy-ion collisions
are needed. The study of kaon properties in dense matter
is also relevant for astrophysics problems, such as the properties
of neutron stars {\cite{prak}}. 

There have been some studies on strangeness production and flow
in heavy-ion collisions at SIS energies 
{\cite{fang,maru,aich,likoli,liko,likofa,cassing}}.
In this contribution we will concentrate on recent development
concerning strangeness production in Ni+Ni collisions at 1-2 AGeV.
We will show that these observables are sensitive to kaon 
properties in dense matter and can thus provide useful information
for the development of chiral perturbation theory in matter and 
for the study of neutron star properties. In Section II, we 
review briefly the current understanding of kaon properties in
nuclear matter, the relativistic transport model with strangeness,
and the various elementary processes for strangeness production.
Our results for strangeness spectra and flow will be reported
in Section III, where we will also compare our results for proton
and pion spectra to experimental data. Finally a brief 
summary is given in Section IV.

\section{THE RELATIVISTIC TRANSPORT MODEL WITH STRANGENESS}

Our study is based on the relativistic transport model 
extended to include the strange degrees of freedom {\cite{koli}}.
The Lagrangian we use is given by
\begin{eqnarray}
{\cal L}&=&{\bar N}(i\gamma^\mu\partial_\mu-m_N+g_\sigma\sigma)N
-g_\omega\bar N\gamma^\mu N\omega_\mu
+{\cal L}_0(\sigma,\omega_\mu)\nonumber\\
&+&{\bar Y}(i\gamma^\mu\partial_\mu-m_Y
+(2/3)g_\sigma\sigma)Y-(2/3)g_\omega\bar Y\gamma^\mu Y\omega_\mu\nonumber\\
&+&\partial^\mu{\bar K}\partial_\mu K
-(m_K^2-{\Sigma_{KN}\over f^2}{\bar N}N){\bar K}K
-{3i\over 8f^2}{\bar N}\gamma^0 N
\bar K \buildrel \leftrightarrow\over \partial_t K.
\end{eqnarray}
In the above the first line gives the usual non-linear $\sigma$-$\omega$
model with the self-interaction of the scalar field. The second
line is for the hyperons which couple to the scalar and vector
fields with 2/3 of the nucleon strength, as in the constituent
quark model. The last line is for the kaon which is derived from
the SU(3) chiral Lagrangian. Note that since the hyperon and 
kaon densities in heavy-ion collisions at SIS energies are 
very small, their contributions to the scalar and vector fields
are neglected. 

From this Lagrangian we get the kaon and antikaon in-medium energies
\begin{eqnarray}
\omega _K=\left[m_K^2+{\bf k}^2-a_K\rho_S
+(b_K \rho_N )^2\right]^{1/2} + b_K \rho_N 
\end{eqnarray}
\begin{eqnarray}
\omega _{\bar K}=\left[m_K^2+{\bf k}^2-a_{\bar K}\rho_S
+(b_K \rho_N )^2\right]^{1/2} - b_K \rho_N 
\end{eqnarray}
where $b_K=3/(8f_\pi^2)\approx 0.333$ GeVfm$^3$, 
$a_K$ and $a_{\bar K}$  are two parameters
that determine the strength of attractive scalar potential for
kaon and antikaon, respectively. If one considers only the
Kaplan-Nelson term, then $a_K=a_{\bar K}=\Sigma _{KN}/f_\pi ^2$.
In the same order, there is also the range term which acts differently
on kaon and antikaon, and leads to different scalar attractions.
We take the point of view that they can be treated as free parameters
and try to constrain them from the experimental observables in
heavy-ion collisions.

In chiral perturbation theory, in the same order as the
Kaplan-Nelson term  there is the range term, which can be
taken into account by renormalizing {\cite{lee}}
\begin{eqnarray}
\Sigma _{KN} \longrightarrow (1-0.37{\omega^2_{K,{\bar K}}\over m_K^2})
\Sigma _{KN}.
\end{eqnarray}
Since $\omega _K\sim m_K$ for the densitis considered here, 
there is a reduction of $\sim 0.63$ from the range term. However,
$f_\pi$ is also density dependent {\cite{br96}},
\begin{eqnarray}
{f_\pi^{*2}(\rho_0)\over f_\pi^2} \approx 0.6.
\end{eqnarray}
We do not know quantitatively the behaviour of $f_\pi^*/f_\pi$ at higher
densities, although we would expect it to continue to decrease with
density. We see that, at least in the range of densities
$\rho \sim \rho_0$, there is considerable cancellation between
effects of the range term and the decrease in $f_\pi$. We therefore
neglect both for $K^+$. 

In the case of $K^-$ meson, the range term drops rapidly with density
as $(\omega _{\bar K}/m_K)^2$ decreases. We thus neglect the range 
term and relate $a_{\bar K}$ to $a_K$, namely,  
$a_{\bar K} \approx  (0.63)^{-1}a_K$.
Note that $f_\pi^2$ in $b_K$ is not scaled with density.
The $b_K$ represents the vector interaction, which is 1/3
of the vector mean field acting on a nucleon. In our
recent paper {\cite{li97}}, it is shown that for $\rho \sim
(2-3)\rho _0$, the region of densities important for
kaon and antikaon production and flow, the nucleon vector 
mean field is estimated to be only about 15\% larger than
$3b_K=9/(8f_\pi^2)$. The increase from scaling $f_\pi^{-2}$ is evidently
largely canceled by the decrease from short-range correlations.
Our parameterization thus incorporates roughly what we know 
about the scaling of $f_\pi$ and of effects from short-range 
interactions.  

Using $a_K=0.22$ GeV$^2$fm$^3$ and $a_{\bar K}=0.35$ GeV$^2$fm$^3$,
the kaon and antikaon effective mass, defined as their energies at
zero momentum, are shown in Fig. 1. It is seen that the kaon mass
increases slightly with density, resulting from near cancellation of
the attractive scalar and repulsive vector potential. The mass
of antikaon drops substantially. At normal nuclear matter density
$\rho _0\approx 0.16$ fm$^{-3}$, the kaon mass increases about
4\%, in rough agreement with the prediction of impulse approximation
based on $KN$ scattering length. The antikaon mass drops by about 22\%,
which is somewhat smaller than what has been inferred from
the kaonic atom data {\cite{gal}}, namely, an attractive
$K^-$ potential of $200\pm 20$ MeV at $\rho_0$.

From Fig. 1 we find that $m_{\bar K} (3\rho_0) \approx 170$ MeV.
The correction for neutron rich matetr in using $m_{\bar K}$
in neutron stars is about 50 MeV upwards, giving 
$m_{\bar K}^* \approx 220$ MeV {\cite{brown}}.
With the electron chemical potential of $\mu _e (3\rho_0)
=214 $ MeV {\cite{thor}}, this implies that kaon condensation
will take place at dnsity $\rho \sim \rho_0$.
 
From the Lagrangian we can also derive equations of motion for
nucleons {\cite{liko}}
\begin{eqnarray}
{d{\bf x}\over dt}= {{\bf p}^*\over \sqrt{{\bf p}^{*2}+m_N^{*2}}},
~~{d{\bf p} \over dt}= - \nabla _x (\sqrt{{\bf p}^{*2}+m_N^{*2}}
+U_V),
\end{eqnarray}
for hyperons
\begin{eqnarray}
{d{\bf x}\over dt}= {{\bf p}^*\over \sqrt{{\bf p}^{*2}+m_Y^{*2}}},
~~{d{\bf p} \over dt}= - \nabla _x (\sqrt{{\bf p}^{*2}+m_Y^{*2}}
+(2/3)U_V),
\end{eqnarray}
and for kaons
\begin{eqnarray}
{d{\bf x}\over dt}= {{\bf p}^*\over \omega _{K,{\bar K}}\mp b_k\rho _N},
~~{d{\bf p} \over dt}= - \nabla _x \omega _{K,{\bar K}},
\end{eqnarray}
where $m_N^*=m_N-U_S$ and $m_Y^*=m_Y-(2/3)U_S$, with
$U_S$ and $U_V$ being nucleon scalar and vector potentials.
The minus and the plus sign in the last equation 
correspond to kaon and antikaon, respectively.

In addition to propagations in their mean field potentials, 
we include typical two-body scattering processes such as 
$BB\leftrightarrow BB$, $NN\leftrightarrow
N\Delta$ and $\Delta\leftrightarrow N\pi$. Kaons and hyperons
are mainly produced from the following baryon-baryon and
pion-nucleon collisions, namely $BB\rightarrow BYK$, 
$\pi N\rightarrow YK$. The cross section for the former is
taken to be the Randrup-Ko parameterization {\cite{ran}},
and the cross section for the latter is taken from Cugnon
{\it et al}. {\cite{cug}}.   
The antikaon production cross section in baryon-baryon 
collisions is taken from the recent work of Sibirtsev 
{\it et al.} {\cite{sib}}, while that in
pion-nucleon collisions is fitted to experimental data.
Antikaons can also be produced
from hyperon-pion collisons through the strangeness exchange process,
namely, $\pi Y\rightarrow {\bar K}N$. The cross section for
this process is obtained from the reverse one, ${\bar K}N\rightarrow 
\pi Y$, by the detailed-balance relation. The latter cross section,
together with the ${\bar K}N$ elastic and absorption cross
sections, are parameterized based on the available experimental
data. We note that the antikaon absorption
cross section is relatively large, and increases significantly
with descreasing antikaon momentum, while kaons and hyperons
undergo mainly elastic scattering with nucleons.

\section{RESULTS AND DISCUSSIONS}

We have studied strangeness production and flow mainly for
Ni+Ni collisions at 1-2 GeV/nucleon. Our results for
proton and pion transverse mass spectra in central Ni+Ni
collisions at 1.06 AGeV are shown in
Fig. 2. We determine our centrality to be $b\le 2$ fm
in order to compare with the FOPI data {\cite{fopi}}
which correspond to a geometric cross section of 100 mb.
To get free protons we applied a density cut of 
$\rho \le 0.15\rho_0$. It is seen that our model
describes both the proton and pion spectra very well.

In the left panel of Fig. 3 we show our results for $K^+$ kinetic energy 
spectra in Ni+Ni collisions at 1.0 AGeV,
with impact parameter $b\le 8$ fm. The solid histogram gives
the results with kaon medium effects, while the dotted 
histogram is the results without kaon medium effects. The open
circles are the experimental data from the KaoS collaboration
{\cite{kaos}}. It is seen that the results with 
kaon medium effects are in good agreement with the data, while
those without kaon medium effects slightly overestimate
the data. We note that kaon feels a slightly repulsive potential,
thus the inclusion of the kaon medium effects reduces the kaon yield.
The slopes of the kaon spectra in the two cases also differ. With
a repulsive potential, kaons are accelerated during the propagation,
leading to a larger slope parameter as compared to the case without
kaon medium effects.

The results for $K^-$ kinetic energy spectra in Ni+Ni
collisions at 1.8 AGeV are shown in the right panel of Fig. 3.
The solid and dotted histograms are the results with and without
kaon medium effects. It is seen that without medium effects, our
results are about a factor 3-4 below the experimental data.
With the inclusion of the medium effects which reduces the
antikaon production threshold, $K^-$ yield increases by about a factor
of 3 and our results are in good agreement  with the data. This
is similar to the findings of Cassing {\it et al.} {\cite{cassing}}.

The KaoS data show that the $K^-$ yield at 1.8 AGeV agrees roughly
with the $K^+$ yield at 1.0 AGeV. This is a nontrial observation.
At these energies, the Q-values for $NN\rightarrow NK\Lambda$ and 
$NN\rightarrow NNK{\bar K}$ are both -230 MeV.  Near their thresholds,
the cross section for the $K^-$ production is about one order of 
magnitude smaller than that for $K^+$ production. In addition, antikaons
are strongly absorbed in heavy-ion collisions, which should further
reduces the $K^-$ yield. The KaoS results of $K^-/K^+\sim 1$
indicate thus the importance of secondary processes such as $\pi Y\rightarrow
{\bar K}N$ {\cite{ko83}}, and medium effects which 
acts oppositely on the kaon and antikaon production in medium.
 
In addition to particle yield, the collective flow of particles
provide complimentary information about hadron properties in dense matter.
In the left panel of Fig. 4 we compare our results for 
proton flow in Ni+Ni collisions at 1.93 AGeV with the experimental
data from the FOPI collaboration {\cite{fopi}}.
Note that both the data and our results include a transverse
momentum cut of $p_t/m>0.5$.
The agreement with the data is very good. The $\Lambda$ flow
in the same system is shown in the right panel of Fig. 4.
The flow of the primodial $\Lambda$'s, as shown in the
figure by short-dashed curve, is considerably smaller 
than that of protons shown by dotted curve. Inclusions
of $\Lambda N$ rescattering (long-dashed curve) and propagation
in potential (solid curve) increase the $\Lambda$ flow in the
direction of proton flow. Both the FOPI {\cite{fopi}}
and the EOS {\cite{eos}} collaborations found that
the $\Lambda$ flow is very similar to that of protons.

In the left panel of Fig. 5 we show our results for $K^+$ flow in
Ni+Ni collisions at 1.93 AGeV, based on three scenarios 
for kaon potentials in nuclear medium. Without  
potential, kaons flow in the direction of nucleon as 
shown by the dotted curve. Without scalar potential, kaons feel a
strong repulsive potential, and they flow in the 
opposite direction to nucleons as shown by the dashed curve.
With both the scalar and vector potentials, kaons feel 
a weak repulsive potential, and one sees the disappearance 
of kaon flow, which is in good agreement with the experimental
data from the FOPI collaboration {\cite{fopi}}.
In the right panel of Fig. 5 we show $K^-$ flow in
Ni+Ni collsions at 1.93 AGeV and $b=4$ fm. Without
medium effects one sees a clear antikaon antiflow signal,
because of strong absorption of $K^-$ by nucleons.
Including the attractive antikaon potential, those antikaons
that survive the absorption are pulled towards nucleon
and thus show a weak antikaon flow signal. 
The experimental measurement of $K^-$ in heavy-ion
collisions will be very helpful in determining $K^-$
properties in nuclear matter.

\section{SUMMARY}

In summary, we studied strangeness ($K^+$, $K^-$, and $\Lambda$)
production and flow in Ni+Ni collisions at 1-2 AGeV. We based
our study on the relativistic transport model including
the strangeness degrees of freedom. We found that strangeness
spectra and flow are sensitive to the properties of strange
hadrons in nuclear medium. The predictions of the chiral
perturbation theory that the $K^+$ feels a weak repulsive potential
and $K^-$ feels a strong attractive potential are in good agreement
with recent experimental data from FOPI and KaoS collaborations. 

\vskip 1cm

The work of GQL, GEB and CHL were supported in part by 
Department of Energy under Grant No. DE-FG02-88ER40388,
while that of CMK was supported in part by the National 
Science Foundation under Grant No. PHY-9509266.
CHL was also partly supported by Korea Science and Engineering
Foundation.

{99}

\newpage
\begin{figure}
\begin{center}
\epsfig{file=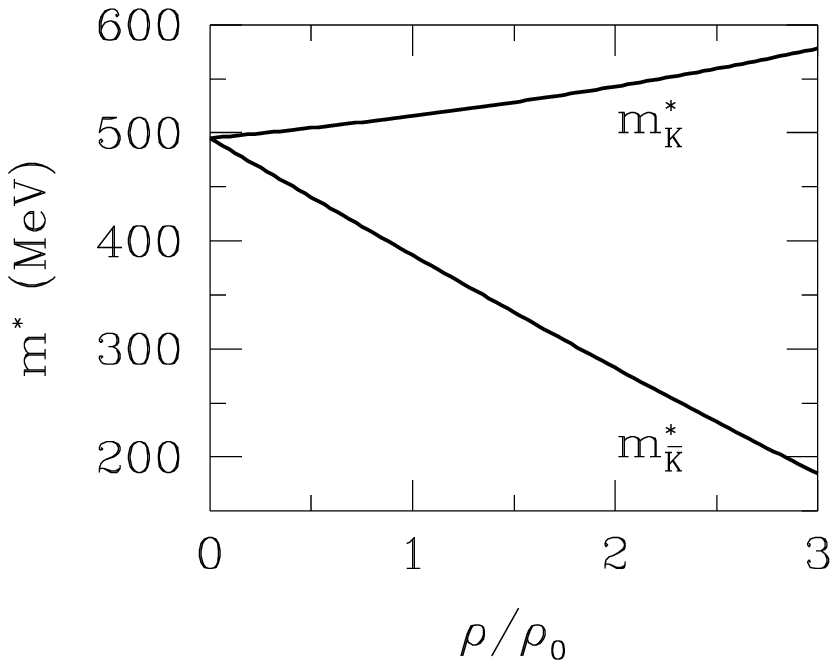,height=7.0in,width=7.0in}
\caption{Kaon and antikaon effective mass as a function of
nuclear density} 
\end{center}
\end{figure}

\newpage
\begin{figure}
\begin{center}
\epsfig{file=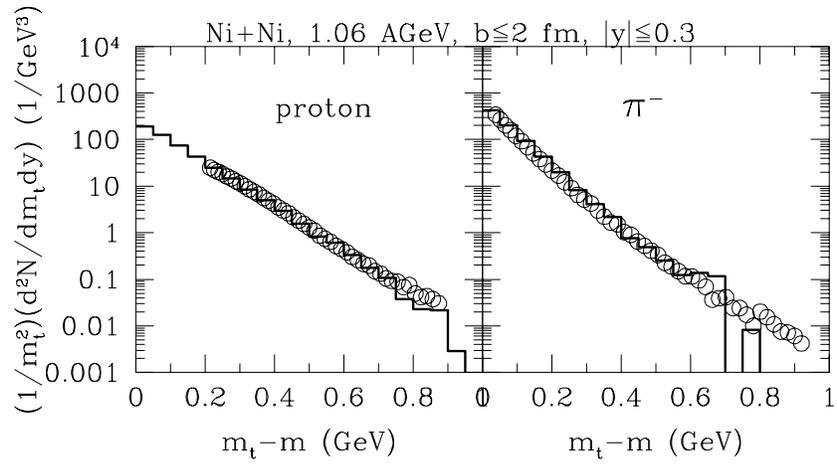,height=7.0in,width=7.0in}
\caption{Proton and pion transverse momentum spectra in
central Ni+Ni collisions at 1.06 AGeV. The open circles
are experimental data from the FOPI collaboration} 
\end{center}
\end{figure}

\newpage
\begin{figure}
\begin{center}
\epsfig{file=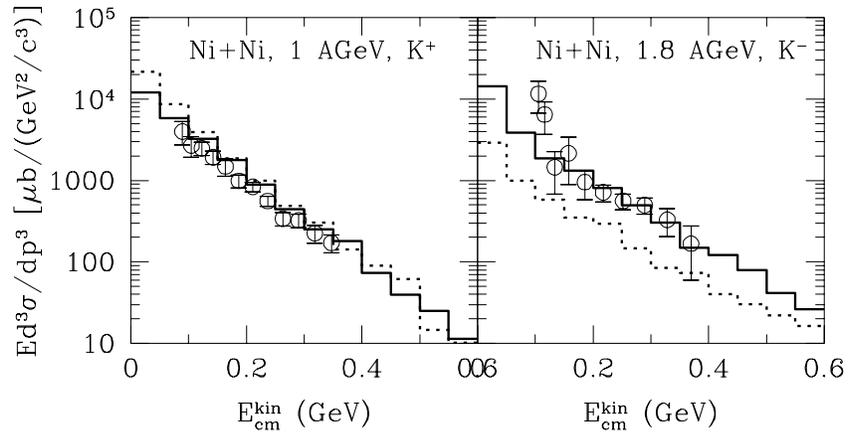,height=7.0in,width=7.0in}
\caption{Left panel: $K^+$ kinetic energy spectra in Ni+Ni collsions at 
1.0 AGeV. Right panel: $K^-$ kinetic energy spectra in Ni+Ni 
collisions at 1.8 AGeV. The solid and dotted histograms are the 
results with and without kaon medium effects.
The open circles are the experimental data 
from the KaoS collaboration.} 
\end{center}
\end{figure}

\newpage
\begin{figure}
\begin{center}
\epsfig{file=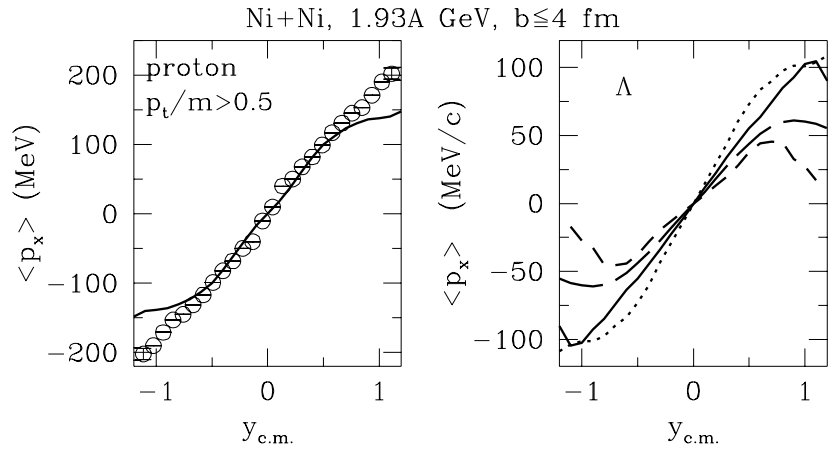,height=7.0in,width=7.0in}
\caption{Left panel: Proton flow in Ni+Ni collisons at 1.93 AGeV and
$b\le 4$ fm, including a transverse momentum cut of $p_t/m>0.5$. 
The open circles are the experimental data from the
FOPI collaboration. Right panel: $\Lambda $ flow.
The short-dashed curve is for primordial $\Lambda$'s, the long-dashed curve
includes elastic $\Lambda N$ scattering, and the solid curve
include both rescattering and propagation. The proton flow
is shown by the dotted curve for comparison.}
\end{center}
\end{figure}

\newpage
\begin{figure}
\begin{center}
\epsfig{file=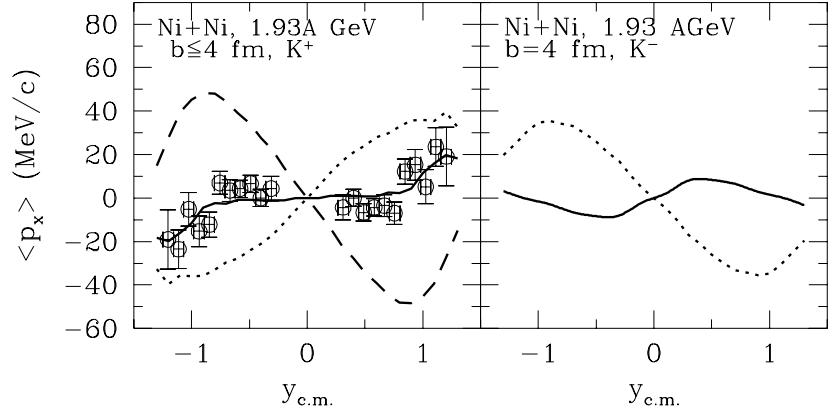,height=7.0in,width=7.0in}
\caption{Left panel: $K^+$ flow in Ni+Ni colliisons at 1.93 AGeV and
$b\le 4$ fm, including a transverse momentum cut of $p_t/m>0.5$. 
The open circles are the experimental data from the
FOPI collaboration.  The dotted curve is the results
without kaon potential, the dashed curve is the results without
kaon scalar potential, and the solid curve gives the results
with both the scalar and vector potential. 
Right panel: $K^-$ flow in Ni+Ni collisions at 1.93 AGeV and $b=4$ fm.
The solid and dashed curves are the results with and without
antikaon medium effects.}
\end{center}
\end{figure}

\end{document}